\documentstyle[prl,aps,preprint,tighten,floats,aps,epsf,psfig]{revtex}
\input epsf
\def\simlt{\stackrel{<}{{}_\sim}}

\def\be{\begin{equation}}
\def\ee{\end{equation}}
\def\bear{\begin{eqnarray}}
\def\eear{\end{eqnarray}}
\def\beqn{\begin{eqnarray}}
\def\eeqn{\end{eqnarray}}

\begin{document}
\draft
\preprint{\vbox{\baselineskip=12pt
\rightline{FERMILAB-Pub-99/136-T}
\vskip0.2truecm
\rightline{hep-ph/9905215}}}

\title{A Resolution to the Supersymmetric CP Problem with Large
Soft Phases via D- branes}
\author{
M. Brhlik${}^{\dagger}$, L. Everett${}^{\dagger}$, G. L.
Kane${}^{\dagger}$,
and J. Lykken${}^*$}
\address{${}^{\dagger}$ Randall Laboratory, Department of Physics\\
          University of Michigan\\
         Ann Arbor, Michigan, 48109, USA \\
${}^*$Theoretical Physics Department\\
Fermi National Accelerator Laboratory\\
Batavia, Illinois, 60510, USA}
\maketitle
\begin{abstract}
We examine the soft supersymmetry breaking parameters that result from
various ways of embedding the Standard Model (SM) on D- branes
within the Type I string picture,
allowing the gaugino masses and $\mu$ to have
large CP- violating phases.  One embedding naturally provides the
relations among soft parameters to satisfy the electron and neutron
electric dipole moment constraints even with large phases, while with 
other embeddings large phases are not allowed. The string
models provide some motivation for large phases in the soft breaking
parameters. The results generally suggest how low energy data might teach
us about Planck scale physics.
\end{abstract}
\vskip2cm
\newpage

The parameters of the Lagrangian of the Minimal Supersymmetric Standard
Model (MSSM) include a number of CP- violating phases, which arise both
in the soft breaking sector and in the phase of the supersymmetric 
higgsino mass parameter $\mu$ (for a careful count see \cite{dimopoulos}).
The presence of these
phases have typically been neglected in phenomenological analyses due to
what traditionally has been called the supersymmetric CP problem:
the electric dipole moments (EDM's) of the fermions receive one-loop
contributions due to superpartner exchange which for large phases can
exceed the experimental bounds. The current bounds for the electron
\cite{eexp} and neutron \cite{nexpnew}
\begin{eqnarray}
|d_e| &<& 4.3 \times 10^{-27} {\ \rm ecm} {\ \rm (95\% \, c.l.)}\\
|d_n| &<& 6.3 \times 10^{-26} {\ \rm ecm} {\ \rm (90\% \, c.l.)}
\end{eqnarray}
were thought to constrain the phases to be
${\cal O}(10^{-2})$ for sparticle masses at the TeV scale
\cite{oldedm,edm,gar}.
However, the results of a recent reinvestigation of
this issue \cite{nath,bgk} 
have demonstrated that cancellations between
different contributions to the electric dipole moments can allow for
phenomenologically viable regions of parameter space
with phases of ${\cal O}(1)$ and light sparticle masses, contrary to
conventional wisdom.  The phases, if nonnegligible,
not only have significant 
phenomenological implications for CP- violating observables (such
as in the K and B systems), but also have important consequences for
the extraction of the MSSM parameters from experimental measurements of 
CP- conserving quantities, since almost none of the Lagrangian
parameters are directly measured \cite{bk}.

The results of \cite{nath,bgk} indicate that the cancellations can
only occur if the soft breaking parameters satisfy certain approximate
relations which are testable in future experiments; such relations may
provide clues to the mechanism of supersymmetry breaking and the form of
the underlying theory. As superstring theory is the only candidate for a
fundamental theory which unifies gravitational and gauge interactions, 
a study of the patterns of CP- violating phases in soft
supersymmetry breaking parameters derived in classes of four-dimensional
superstring models is well motivated.  In particular, we wish to determine
if these models either allow for or predict phenomenologically viable
large phase solutions.  We find that in some models large phases naturally
arise in a manner that leads to results consistent with the EDM
constraints, while in
others the constraints cannot be satisfied.

CP is a discrete gauge symmetry in string theory, and thus can only
be broken spontaneously \cite{dine}. If this breaking occurs via the
dynamics of compactification and/or supersymmetry breaking, then the
four-dimensional effective field theory will exhibit explicit
CP- violating phases.
The origin of the nonperturbative dynamics of supersymmetry breaking
in superstring theory remains unresolved, but progress can be made by
utilizing a phenomenological approach first advocated by Brignole,
Ib\'a\~nez, and Mu\~noz \cite{bim}.
In their approach, it is assumed that supersymmetry breaking effects
are communicated dominantly via the $F$- component vacuum expectation
values (VEV's) of the dilaton $S$ and moduli $T_i$, which are
superfields uncharged under the SM gauge group and are generically
present in four-dimensional string models.  The effects of the unknown
supersymmetry breaking dynamics are then encoded in a convenient
parameterization of these $F$- component VEV's \cite{bim,ibanez}:
\begin{eqnarray}
F^S&=&\sqrt{3}(S+S^{*})m_{3/2}\sin \theta e^{i\alpha_S}\nonumber\\
F^i&=&\sqrt{3}(T_i+T^{*}_i)m_{3/2} \cos \theta \Theta_i e^{i\alpha_i},
\end{eqnarray}
in which $m_{3/2}$ is the gravitino mass and $\theta$, $\Theta_i$ are
Goldstino angles (with $\sum_i \Theta_i^2=1$), which measure the
relative contributions of $S$ and $T_i$ to the supersymmetry breaking.
The $F$- component VEV's are assumed to have arbitrary complex phases
$\alpha_S$, $\alpha_i$, which provide sources for the CP-
violating phases in the soft terms.  

Within particular classes of four-dimensional string models the couplings
of the dilaton, moduli, and MSSM matter fields are calculable (at least at
the tree-level), leading in turn to specific patterns of the soft breaking 
parameters.   We have analyzed the phase structure of the soft terms
arising in three classes of four-dimensional string models: (i) orbifold
compactifications of perturbative heterotic string theory, (ii)
Ho\v{r}ava-Witten type M theory compactifications, and (iii) the Type IIB
orientifold models, which are examples within the general Type I string
picture. In our analysis \cite{bekl}, we take the predictions
for the soft breaking parameters at the string scale, and evolve the
parameters to the electroweak scale using renormalization group equations
(RGE's). Our results indicate that the patterns of CP- violating phases
consistent with the EDM constraints strongly depend on the type of string
model under consideration.  

First, we note that the general results of \cite{bgk} demonstrate that
sufficient cancellations among the various contributions to
the EDM's are difficult to achieve unless there are large relative phases
in the soft masses of the gaugino sector. This feature is due to the
approximate $U(1)_R$ symmetry of the Lagrangian of the MSSM \cite{thomas},
which allows one of the phases of the gaugino masses to be set to zero at
the electroweak scale without loss of generality
\cite{thomas,bgk,wagner,reparameterization}.  
Furthermore, the phases of the gaugino mass parameters do not run at 
one-loop order, and thus at the electroweak scale only deviate from the
string-scale values by small two-loop corrections. Therefore, if the
phases of the gaugino masses are universal at the string scale, they will
be approximately zero at the electroweak scale (after the $U(1)_R$
rotation).  Cancellations among the chargino and neutralino
contributions to the electron EDM are then necessarily due to the
interplay between the phases of $A_e$ and $\mu$ ($\varphi_{A_e}$ and
$\varphi_{\mu}$, respectively). 
The analysis of \cite{bgk} demonstrates that cancellations are then difficult
to achieve  as the pure gaugino part of the neutralino diagram adds
destructively with the contribution from the gaugino-higgsino mixing,
which in turn has to cancel against the chargino
diagram. As a result, the cancellation mechanism is generally
insufficient, and hence in this case the phases of the other soft
breaking parameters as well as the $\mu$ parameter must naturally be
$\simlt  10^{-2}$ (the traditional bound) \cite{edm}.

This feature is predicted \cite{choi} in perturbative heterotic models at
tree level, due to the universal coupling of the dilaton to all gauge
groups in the tree-level gauge kinetic function $f_a=k_aS$ (in which $k_a$ 
is the Ka\v{c}-Moody level of the gauge group).  In these
models, nonuniversal gaugino masses do occur at the loop
level due to moduli-dependent threshold corrections.  Hence, nontrivial
CP effects  require both moduli dominance and large threshold
effects in order to overcome the tendency of the dilaton $F$- term to
enforce universal gaugino masses \cite{bekl,foot1}.
Similar statements apply to the soft breaking parameters derived in the
Ho\v{r}ava-Witten M theory \cite{horavawitten} type scenarios
\cite{munoz,ovrut}. In these scenarios, the gaugino mass parameters are
universal because the observable sector gauge groups all arise from one of
the ten-dimensional boundaries.
Therefore, only a very small fraction of the
$\varphi_{\mu}-\varphi_{A_e}$ parameter
space leads to models allowed by the electron EDM \cite{bekl}.

However, within the more general Type I string picture, 
there is the possibility of nonuniversal gaugino masses at tree level,
which has important implications for the possible CP- violating effects.
We focus on examples within the four-dimensional Type IIB orientifold
models, in which the Type IIB theory is compactified on orientifolds
(which are orbifold compactifications accompanied by the worldsheet parity
projection)\cite{sagnotti,kakushadze,shiu,ibanez}.  In these
models, consistency conditions (tadpole cancellation) require the addition
of open string (Type I) sectors and Dirichlet branes, upon which the open
strings must end. It is important to note that orientifolds are illustrative
of a much larger class of models in the Type I picture, containing
more general configurations of nonperturbative objects (e.g. D- brane
bound states) in more general singular backgrounds (e.g. conifolds
\cite{conif}).  

While the number and type of D- branes required in a given
model depends on the details of the orientifold group, we consider the
general situation with one set of nine-branes  and three
sets of five-branes ($5_i$), in which the index $i$ labels the complex
coordinate of the internal space included in the world-volume of the
five-brane.   Each set of coincident D- branes gives rise to a
(generically non-Abelian) gauge group. Chiral matter fields also arise
from the open string sectors, and can be classified into two categories.
The first category consists of open strings which start and end on D-
branes of the same sector, for which
the corresponding matter fields are charged under the gauge group
(typically in the fundamental or antisymmetric tensor
representations) of that set of branes.  The second category consists
of open strings which start and end on different sets of branes; in this 
case, the states are bifundamental representations under the two gauge
groups from the two D- brane sectors.

Model-building techniques within this framework are
at an early stage and there is as yet no ``standard" model;
furthermore this framework does not provide any generic solution
to the related problems of the runaway dilaton, supersymmetry breaking,
and the cosmological constant. On the other hand, recent
investigations \cite{ibanez} have uncovered the generic
structure of the tree-level couplings of this class of models. The
results indicate that the phenomenological implications of these models
crucially depend on the embedding of the SM gauge group into the different
D- brane sectors, and may have distinctive properties from those of the
perturbative heterotic models traditionally considered in studies of
superstring phenomenology.

Of particular importance for the purposes of this study is that
the dilaton no longer plays a universal role as it did in the
perturbative heterotic case, as can be seen from the form of the
(tree-level) gauge kinetic functions determined in \cite{ibanez}
using T-duality and the form of the Type I low energy effective
action:
\begin{eqnarray}
\label{orientifoldf}
f_{9}&=&S\nonumber\\
f_{5_i}&=&T_i.
\end{eqnarray}
This result illustrates a distinctive feature of this class of
models, which is that in a sense there is a different ``dilaton" for each
type of brane. This fact has important implications both for gauge
coupling unification \cite{ibanez} and the patterns of gaugino masses
obtained in this class of models, which strongly depend on the details of
the SM embedding into the five-brane and nine-brane sectors.
For example, in the case in which the SM gauge group is associated with a
single D- brane sector, the pattern of the gaugino masses resembles that
of the tree-level gaugino masses in the weakly coupled heterotic models
\cite{bekl}, as can be seen from the similarity between
(\ref{orientifoldf}) and the corresponding tree-level expression for $f$
in the perturbative heterotic case.  

However, an alternate possibility is that the SM gauge group is not
associated with a single set of branes, but rather is embedded within
multiple D- brane sectors.   We consider the case in which 
$SU(3)$ and $SU(2)$ originate from the $5_1$ and $5_2$ sectors,
respectively \cite{foot3}.  In this case, the quark doublet states 
necessarily arise from open strings connecting the two D- brane sectors; 
as these states have a nontrivial hypercharge assignment, their presence 
restricts $U(1)_Y$ to originate from the $5_1$ and/or $5_2$ sectors as well.  
We consider two models of the resulting soft terms corresponding to the
two simplest possibilities for the hypercharge embedding, which are to
have $U(1)_Y$ in either the $5_1$ or $5_2$ sector. 
Depending on the details of the hypercharge embedding, the remaining MSSM
states may either be states which (in analogy with the quark
doublets) are trapped on the intersection of these two sets of
branes, or states associated with the single $5_i$ sector which contains
$U(1)_Y$. In any event the natural starting point for constructing models
with these features are orientifolds which realize identical GUT gauge
groups and massless matter on two sets of intersecting 5- branes.
The existence of such symmetrical arrangements is often guaranteed by
T-duality. For example,
Shiu and Tye \cite{shiu}
have exhibited an explicit model which realizes the Pati-Salam gauge
fields of
$SU(4)$$\times$$SU(2)_L$$\times$$SU(2)_R$ and identical chiral matter
content on two sets of 5- branes.
Additional Higgsing and modding by discrete symmetries
could then in principle produce the asymmetrical structures
outlined above.

In the case with $U(1)_Y$ and $SU(3)$ from
the $5_1$ sector, the gaugino masses and $A$ terms take the form (see the
general formulae in \cite{ibanez}):
\begin{eqnarray}
\label{model3}
M_{1}&=&\sqrt{3}m_{3/2}\cos \theta \Theta_1
e^{-i\alpha_1}=M_3=-A_{t,e,u,d}\nonumber\\
M_{2}&=&\sqrt{3}m_{3/2}\cos \theta \Theta_2 e^{-i\alpha_2}.
\end{eqnarray}
Note that in this case the phases $\varphi_1$ and $\varphi_3$ of the
mass parameters $M_1$ and $M_3$ are equal and distinct from that of the
$SU(2)$ gaugino mass parameter $M_2$. The soft mass-squares are given by
\begin{eqnarray}
\label{model4}
m^2_{5_15_2}&=&m^2_{3/2}(1-\frac{3}{2}(\sin^2 \theta+\cos^2 \theta
\Theta_3^2) ) \nonumber\\
m^2_{5_1}&=&m^2_{3/2}(1-3\sin^2 \theta ).
\end{eqnarray}
The $SU(2)$ doublets of the MSSM are necessarily states 
which arise from open strings which connect the $5_1$ and $5_2$ brane
sectors, and hence have mass-squares given by the above expression for
$m^2_{5_15_2}$.  The $SU(2)$ singlets may either be fields of the same
type or states which originate from the $5_1$ sector, although we note
that requiring the presence of the MSSM Yukawa couplings indicates that 
these states should be of the latter type\cite{ibanez}.
Similar expressions apply for the case in which
$U(1)_Y$ and $SU(2)$ are associated with the same five-brane sector,
although in this case the relations among the phases are
$\varphi_1=\varphi_2\neq \varphi_3$.  In each of these models, $\mu$ and
$B$ are in principle free complex parameters in the analysis (although
their phases are related by the approximate PQ symmetry of the MSSM
Lagrangian; see \cite{thomas,bgk,bekl} for further details).

In our numerical analysis of these models, we impose the boundary
conditions (\ref{model3}) and (\ref{model4}) 
at the GUT scale $M_{G}=3 \times 10^{16}$ GeV (where we
assume the couplings unify), and evolve the parameters to the electroweak
scale via the renormalization group equations
\cite{foot4}.  The sparticle masses and the CP- violating phases
depend on the free parameters $m_{3/2}$, $\theta$, $\Theta_i,\ i=1,2,3$,
which are related by $\Theta_1^2+\Theta_2^2+\Theta_3^2=1$, as well as the
two phases $\alpha_1$ and $\alpha_2$ (physical results only depend on
$\alpha_1-\alpha_2$). To avoid negative scalar mass-squares we restrict
our consideration to values of $\theta$ which satisfy $\sin^2\theta
<\frac{1}{3}$, and also assume that $\Theta_3=0$ (indicating that the
modulus $T_3$ associated with the $5_3$ brane sector plays no role in
supersymmetry breaking, and thus is essentially decoupled from the
observable sector).
$B$ and $\mu$ are in principle free parameters, as they are not
determined by this embedding. However, we explore the phenomenologically
motivated scenario in which the electroweak symmetry is broken radiatively
as a result of RGE evolution of the Higgs masses $m_{H_1}^2$ and
$m_{H_2}^2$. As the minimization conditions are imposed at
the electroweak scale, the values of $B\mu$ and $|\mu |^2$ can be
expressed in  terms of $\tan\beta$ and $M_Z$ \cite{rad}. We note that 
even under these assumptions $\varphi_{\mu}$ remains an independent
parameter, and thus the model depends on two phases at the GUT scale:
$\alpha_1-\alpha_2$ and $\varphi_{\mu}$ (though due to RG running the
phases of the $A$ terms at the electroweak scale will deviate from
their string-scale values).  Also, the squark and slepton masses from
(\ref{model4}) are of the same order as the gaugino masses, so they would
allow very large EDM's
if there were no relations among the soft masses and phases.

We find, remarkably, that in order to satisfy the experimental
constraints on the electron and neutron
EDM's in this model, the large individual contributions from chargino,
neutralino and gluino loops do not have to be suppressed by small CP
phases.
A cancellation between the chargino and neutralino loop
contributions naturally causes the electron EDM to be acceptably small. As
emphasized in \cite{bgk}, the contributions to chargino and neutralino
diagrams from gaugino-higgsino mixing naturally have opposite signs and
the additional $\varphi_1$ dependence of the neutralino exchange
contribution can provide
for a match in size between the chargino and neutralino contributions.
In the neutron case, the contribution of the
chargino loop is offset by the gluino loop contributions to the electric
dipole operator $O_1$ and the chromoelectric dipole operator $O_2$. Since
$\varphi_1=\varphi_3$ in this scenario, the sign of the gluino
contribution is fixed and it automatically has the correct sign to balance
the chargino contribution in the same region of gaugino
phases which ensures cancellation in the electron case. This simple and
effective mechanism therefore provides extensive regions of parameter
space where the electron and neutron EDM constraints are satisfied
simultaneously while allowing for ${\cal O}(1)$ CP- violating phases.

\vskip 2mm

\begin{figure}[h!]
\centering
\epsfxsize=6in
\hspace*{0in} 
\epsffile{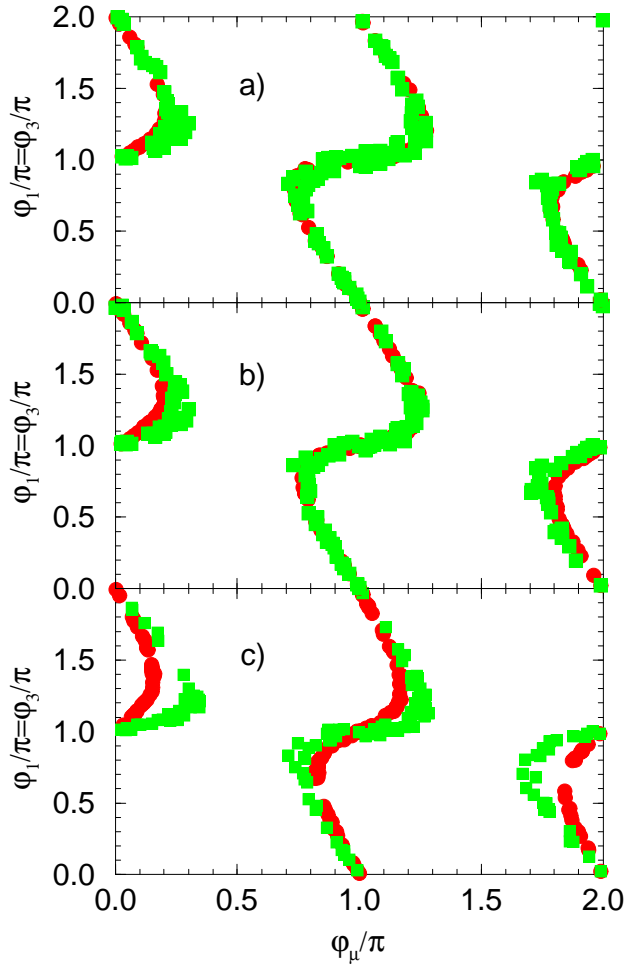}
\bigskip
\caption{Illustration of the overlap between the regions allowed by the
electron and neutron EDM constraints. We choose $m_{3/2}=150\,
\rm{GeV}$, $\theta=0.2$ and $\tan\beta=2$, and assume that the EW symmetry
is broken radiatively. Allowed points are shown for {\it a)}
$\Theta_1=0.85$,  {\it b)} $\Theta_1=\sqrt{\frac{1}{2}}$, and {\it c)}
$\Theta_1=0.55$. The black circles are the points allowed for the eEDM and
grey blocks for the nEDM.  In {\it a)}, the overlap is essentially total
so only the grey blocks show up; in {\it b)} and {\it c)} the overlap
becomes less good. The sensitivity is encouraging, suggesting that the
experimental data can determine the Goldstino angles.} 
\label{figone} 
\end{figure}

In order to demonstrate the coincidence of the regions allowed by the 
experimental constraints on the EDM's, we consider a specific set  of
relevant parameters. 
We choose $m_{3/2}=150\,\rm{GeV}$, $\theta=0.2$ and $\tan\beta=2$ which
leads to a reasonably light spectrum of the superpartners and require
that the EW symmetry is to be broken radiatively. In Fig. 1, we plot the
allowed regions for both electron and neutron EDM depending on the
values of $\Theta_1$ and $\Theta_2=\sqrt{1-\Theta_1^2}$ while $\Theta_3$
is set to zero. Frame {\it a)}, where $\Theta_1=0.85$, shows a very
precise  overlap between the electron and neutron EDM allowed regions 
resulting from the the fact that $\varphi_1=\varphi_3$ and the
cancellation  mechanism works similarly in both cases. In frame {\it b)}, 
we consider the situation in which the magnitudes of all three gaugino
masses are unified in magnitude while still allowing for different
phases due to different origin of $M_1$ and $M_3$ compared to $M_2$.
Finally, in frame {\it c)} we set  $\Theta_1=0.55$ and study the
situation when the magnitude of $M_2$ is significantly larger than 
that of $M_1=M_3$. We find that in this scenario the alignment between
the EDM allowed regions is spoiled and only small CP- violating phases
are allowed by the experimental constraints.
The behavior illustrated in Fig. 1 for a particular choice of 
parameters is quite general provided the gaugino masses at the
unification scale are close in magnitude. We considered only
small and moderate values of $\tan\beta$ since for models with 
large $\tan\beta$ new types of contributions can become important 
\cite{pilaftsis}.      

\begin{figure}[h!]
\centering
\epsfxsize=5.25in
\hspace*{0in} 
\epsffile{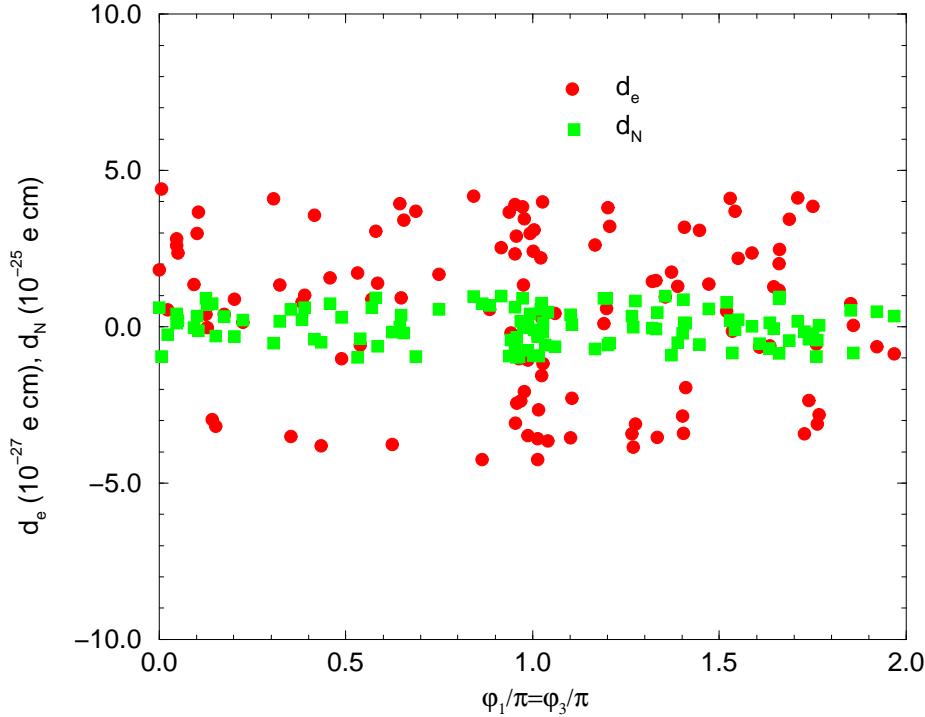}
\bigskip
\caption{Range of the electron and neutron EDM values vs. 
$\varphi_1=\varphi_3$ predicted by Eqs. (\ref{model3}) and (\ref{model4})
for the parameters of Figure 1{\it a)}.  All of the points are allowed by
the experimental bounds on the EDM's (note the different scales for the
eEDM and nEDM).}
\label{figtwo}
\end{figure}

The cancellation mechanism in this scenario provides a remarkably large
range of allowed CP- violating soft phases and requires a specific 
correlation between $\varphi_{\mu}$ and  $\varphi_1=\varphi_3$ as shown
in Fig. 1. It is also interesting to observe that the actual values of
the electron and neutron EDM's for the allowed  points in the phase
parameter space are typically slightly below the experimental limit and
should be within the reach of the next generation of EDM measuring 
experiments. In Fig. 2 we plot the EDM values for the allowed points in
the case of $\Theta_1=0.85$ with all the other parameters set to the
same values as in previous discussion of Fig. 1. This indicates that 
if the  CP- violating phases indeed originate from this type of 
D- brane configuration, non-zero measured values for both EDM's 
much bigger than the SM prediction can be expected.

Equally remarkably, if we modify the way the SM is embedded in the D-
brane sectors we are unable to satisfy the EDM constraints with
nonnegligible phases. For example, the other possibility of arranging the
SM gauge groups, such that $U(1)_Y$ is instead on the $5_2$ brane with
$SU(2)$, does not allow for large phase solutions.  
The reasons for this behavior are similar to that of the
Ho\v{r}ava-Witten scenario: we can use the $U(1)_R$ symmetry of the soft
terms to put $\varphi_2=\varphi_1=0$, which severely limits the
possibility of cancellation between the chargino and neutralino
contributions to the electron EDM. The effect of $\varphi_{A_e}$ alone is
not enough to offset the potentially large chargino contribution and only
a very narrow range of values of $\varphi_{\mu}$ (close to $0$,
$\pi$,$\ldots$) passes the electron EDM constraint.

There are a number of interesting implications of these results:
\begin{itemize}

\item They show explicitly how relations among soft parameters such as
Eqs. (\ref{model3}) and (\ref{model4}) can naturally give small EDM's even
with large phases.

\item They illustrate how we are able to learn about (even nonperturbative)
Planck scale physics using low energy data.  If the soft phases are
measured in (say) collider superpartner data, or at B factories, and
found to be large, we have seen that they may provide guidance as to how
the SM is to be embedded on branes.

\item They illustrate very simply that large soft phases are at least
consistent with, and perhaps motivated by, some string models. In
particular, the requirement that the phases (but not necessarily the
magnitudes) of the gaugino mass parameters are nonuniversal for viable
large phase solutions can naturally be realized in Type I models 
in which the SM gauge group is split among different brane sectors.

\item They suggest that $d_n$ and $d_e$ are not much smaller than the
current limits.

\end{itemize}

\acknowledgments
L. E. thanks M. Cveti\v{c} for many helpful discussions and
suggestions, and J. Wang for comments on the manuscript.  This work was 
supported in part by the U. S. Department of Energy contract
DE-AC02-76CH03000.

\def\B#1#2#3{\/ {\bf B#1} (19#2) #3}
\def\NPB#1#2#3{{\it Nucl.\ Phys.}\/ {\bf B#1} (19#2) #3}
\def\PLB#1#2#3{{\it Phys.\ Lett.}\/ {\bf B#1} (19#2) #3}
\def\PRD#1#2#3{{\it Phys.\ Rev.}\/ {\bf D#1} (19#2) #3}
\def\PRL#1#2#3{{\it Phys.\ Rev.\ Lett.}\/ {\bf #1} (19#2) #3}
\def\PRT#1#2#3{{\it Phys.\ Rep.}\/ {\bf#1} (19#2) #3}
\def\MODA#1#2#3{{\it Mod.\ Phys.\ Lett.}\/ {\bf A#1} (19#2) #3}
\def\IJMP#1#2#3{{\it Int.\ J.\ Mod.\ Phys.}\/ {\bf A#1} (19#2) #3}
\def\nuvc#1#2#3{{\it Nuovo Cimento}\/ {\bf #1A} (#2) #3}
\def\RPP#1#2#3{{\it Rept.\ Prog.\ Phys.}\/ {\bf #1} (19#2) #3}
\def\etal{{\it et al\/}}

\bibliographystyle{prsty}

\begin{thebibliography}{99}
\bibitem{foot1}{The corresponding analysis of the dilaton-dominated scenario
was recently reported in \cite{barrkhalil}; their results demonstrate
this behavior explicitly.  We refer the reader also to \cite{bailin} and 
references therein for the analysis of the $CP$-violating phases in the
gaugino condensation approach.}
\bibitem{foot3}{We choose the $5_i$ sectors for the sake of
definiteness.  We could also have assumed that one of the gauge groups
arises from the nine-brane sector and obtain similar results for the $CP$-
violating effects. However, in this case \cite{ibanez} it may be more
difficult to obtain consistent unification of the gauge couplings at the
GUT scale (although this point is not crucial for the purposes of this
study).}
\bibitem{foot4}{For the sake of simplicity, we utilize the freedom in Type I
string models to choose the string scale to coincide with the GUT scale;
this possibility is not valid in the perturbative heterotic case, which
fixes $M_{String} \sim 5 \times 10^{17}\, \rm{GeV}$ \cite{kaplunovsky}. It
is beyond the scope of this paper to consider all possibilities, and we
refer the reader to a comprehensive discussion of
this issue and its implications for gauge coupling unification in
\cite{ibanez}.}
\end{thebibliography}

\end{document}